\documentclass[
 reprint,
 amsmath,
 amssymb,
 superscriptaddress,
 aps,
 prl,
 groupedaddress
]{revtex4-1}

\usepackage{graphicx}
\usepackage{bm}
\usepackage{gensymb}
\usepackage[utf8]{inputenc}

\usepackage{array,mathtools,amssymb,booktabs}
\newcolumntype{C}{>{$}c<{$}}
\AtBeginDocument{
\heavyrulewidth=.08em
\lightrulewidth=.05em
\cmidrulewidth=.03em
\belowrulesep=.65ex
\belowbottomsep=0pt
\aboverulesep=.4ex
\abovetopsep=0pt
\cmidrulesep=\doublerulesep
\cmidrulekern=.5em
\defaultaddspace=.5em
}

\usepackage{hyperref}
\usepackage{siunitx}
\begin{document}


\title{Laser Cooling of Radium Ions}

\author{M. Fan}
\author{C. A. Holliman}
\author{A. L. Wang} 
\author{A. M. Jayich}
\email{jayich@gmail.com}
\affiliation{Department of Physics, University of California, Santa Barbara, California 93106, USA}
\affiliation{California Institute for Quantum Entanglement, Santa Barbara, California 93106, USA}

\date{\today}

\begin{abstract}
The unstable radium nucleus is appealing for probing new physics due to its high mass, octupole deformation and energy level structure.  Ion traps, with long hold times and low particle numbers, are excellent for work with radioactive species, such as radium and radium-based molecular ions, where low activity, and hence low total numbers, is desirable.   We address the challenges associated with the lack of stable isotopes in a tabletop experiment with a low-activity ($\sim 10 \ \mu\mathrm{Ci}$) source where we laser-cool trapped radium ions. With a laser-cooled radium ion we measured the $7p\ ^2P_{1/2}^o$ state's branching fractions to the ground state, $7s\ ^2S_{1/2}$, and a metastable excited state, $6d\ ^2D_{3/2}$, to be $p=0.9104(7)$ and $0.0896(7)$, respectively. With a nearby tellurium reference line we measured the $7s\ ^2S_{1/2} \rightarrow 7p\ ^2P_{1/2}^o$ transition frequency, \num{640.09663}(6) THz.

\end{abstract}

\maketitle

Radium, the heaviest alkaline earth element, has no stable isotopes.  Singly ionized radium's simple electronic structure is amenable to optical pumping and laser cooling with wavelengths far from the challenging UV of most alkaline earth type ions.  Radium's heavy nucleus, atomic number $Z=88$, is well suited to searches for new physics, where sensitivity to symmetry breaking forces scales $\propto Z^3$  \cite{Bouchiat1997, Commins2007}.  Certain radium isotopes, such as radium-225, have additional nuclear structure enhancements to $CP$ (charge-parity) violating new physics \cite{Auerbach1996, Safronova2018}.  Setting limits to sources of $CP$ violation will help us understand the baryon asymmetry in the observed Universe \cite{Dine2003}.

Pioneering work with trapped $\mathrm{HfF}^+$ molecular ions has made significant progress in constraining leptonic $CP$ violation, and has rigorously studied potential systematic effects for future experiments \cite{Cairncross2017}.  A complementary hadronic $CP$ violation experiment with radioactive molecular ions $\mathrm{RaOH}^+$, or $\mathrm{RaCOH}_3^+$ \cite{Kozyryev2017} is an intriguing possibility, where the low densities and long hold times of ion traps are well matched to working with radioactive isotopes, because low total activity is desirable. The radium-225 nucleus ($I=1/2$)  has octupole deformed parity doublets that enhance sensitivity to $CP$ violating forces by a factor of 100-1000 compared to the current touchstone atomic system, $^{199}\mathrm{Hg}$ \cite{Dobaczewski2005, Graner2016, Parker2015}.  A radium-based molecular ion, such as $^{225}\mathrm{RaOH}^+$, has an additional sensitivity advantage because of the molecule's closely spaced, opposite parity electronic states in addition to the enhancements from the closely spaced, opposite parity radium nuclear states.  Trapped and laser-cooled radium ions could be the starting point for generating such radium-based molecular ions, where optical pumping $\mathrm{Ra}^+$ may provide control of chemical reactions to produce $\mathrm{RaOH}^+$, as seen in other alkaline earth ions $\mathrm{Ca}^+$ and $\mathrm{Be}^+$ \cite{Okada2003, Yang2018}. 

A single laser-cooled radium ion is also a candidate for atomic parity nonconservation (PNC) measurements, as the massive radium nucleus enhances PNC effects and the simple electronic structure is appealing for the requisite calculations \cite{Fortson1993, NunezPortela2014}.  There are many radium isotopes, including several that were previously trapped \cite{Versolato2010}, which can further reduce atomic and nuclear structure uncertainty by measuring across a chain of isotopes \cite{Antypas2018}.  Such PNC measurements could help our understanding of neutron matter, or potentially uncover new physics \cite{Fortson1990, Brown2009, Davoudiasl2014, Safronova2018}.

\begin{figure}
    \centering
    \includegraphics[scale=1.0]{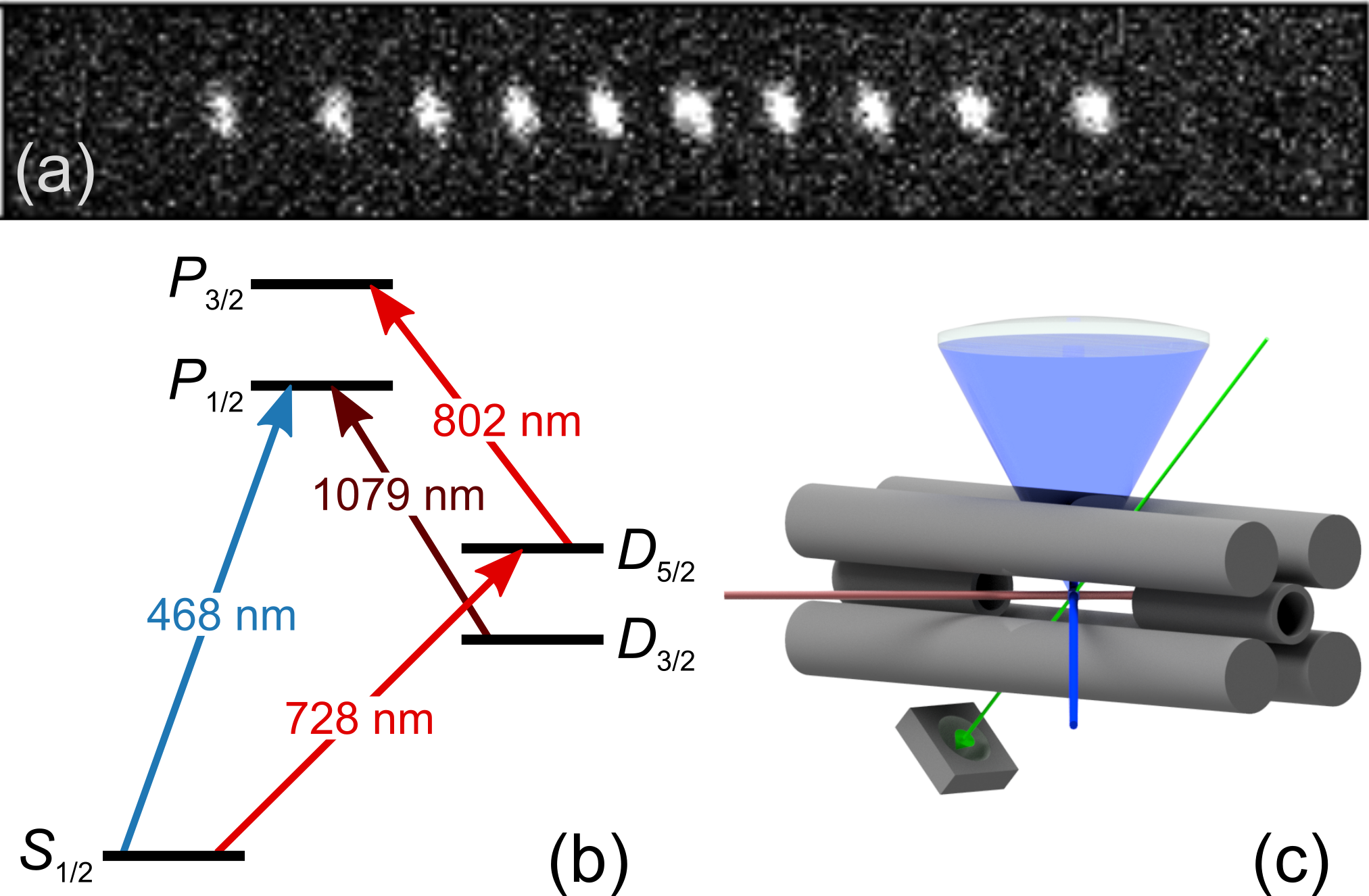}
    \caption{A Coulomb crystal of 10 trapped radium-226 ions (a) that were loaded into the trap via laser ablation and laser cooled with a combination of 468 nm and 1079 nm radiation. The relevant level structure of $\mathrm{Ra}^+$ for the laser cooling and measurements done in this work are shown in (b), in addition to the transitions necessary for controlling the ion via the narrow $^2S_{1/2} \rightarrow {}^2D_{5/2}$ quadrupole transition at 728 nm and the $^2D_{5/2} \rightarrow {}^2 P_{3/2}^o$ dipole transition at 802 nm.  In (c) the ion trap is depicted with the radium ablation target and the ablation laser (green), the 468 nm cooling light (blue), and the 1079 nm repump light (red).  The diagonal rf electrodes are separated by 6 mm and the end cap electrodes by 15 mm.  The rf trapping frequency is 2.1 MHz.}
    \label{fig:trap_levels_and_chain}
\end{figure}

For quantum simulation with trapped ions, qubit states protected from environmental noise with long lifetimes are favorable.  A spin-$1/2$ nucleus, such as in $^{171}\mathrm{Yb}^+$, $^{133}\mathrm{Ba}^+$ \cite{Hucul2017}, or $^{225}\mathrm{Ra}^+$, provides such levels that are first-order insensitive to magnetic fields.  The state is typically read out through optical cycling, with readout fidelity limited by the $P_{1/2}$ state's hyperfine splitting (2.1 GHz in $^{171}\mathrm{Yb}^+$).  Though a massive nucleus is at odds with high secular frequencies, it is desirable for its large hyperfine interactions, as off-resonant pumping during qubit readout decreases quadratically with hyperfine splitting.  The $P_{1/2}$ hyperfine splitting of $^{225}\mathrm{Ra}^+$ is 5.4 GHz \cite{Wendt1987}, which suppresses the qubit readout error by a factor of $\sim$ 8 compared to $^{171}\mathrm{Yb}^+$ \cite{Noek2013}.  Radium also has favorable transitions where abundant optical power and photonic technology are available, see Fig. \ref{fig:trap_levels_and_chain} (b).  The radium ion supports optical qubits on the $S_{1/2} \rightarrow D_{5/2}$ transition.  The $D_{5/2}$ state of $^{225}\mathrm{Ra}^+$, like the ground state, has 2 hyperfine ``clock'' states, which, when combined with the ground state qubit levels, offers the possibility to simulate spin-1 or spin-$3/2$ physics with four magnetic field insensitive states \cite{Senko2015}.

In this work we trap and laser cool $^{226}\mathrm{Ra}^+$ ($I=0$) to form Coulomb crystals, as seen in Fig. \ref{fig:trap_levels_and_chain} (a).  We used the crystallized radium ions to measure the $7p\ ^2P_{1/2}^o$ state's branching fractions to the $7s\ ^2S_{1/2}$ and $6d\ ^2D_{3/2}$ states, a necessary measurement to determine dipole matrix elements for the respective transitions.  Our measurement at 2 digits of precision is sufficient to support optical pumping or basic simulations with optical Bloch equations, but we extended the measurement to higher precision to support PNC measurements in $\mathrm{Ra}^+$ at the $0.8\%$ level \cite{Fortson1993, Wieman2019}.   We also measure the $7s\ ^2S_{1/2} \rightarrow 7p\ ^2P_{1/2}^o$ transition frequency with respect to a $\mathrm{Te}_2$ molecular absorption line, which establishes a convenient frequency reference for the radium-226 ion's most important transition \cite{Cariou1980a}.

In previous work at a nuclear facility singly ionized radium isotopes 209 through 214 were produced and trapped \cite{Traykov2008, NunezPortela2014}.  We apply a different technique to trap radium-226. The radium is ionized and loaded into the trap by ablation with a 532 nm $\sim10$ mJ  pulse from a Nd:YAG laser with 0.5(1) mm $1/e$ intensity diameter.  The ion trap's rf trapping voltage is switched on 20 $\mu \mathrm{s}$ after the ablation pulse to enhance radium ion loading efficiency \cite{Schneider2016}.  The radium was received as $^{226}\mathrm{RaCl}_2$ in 5 mL 0.1 M HCl solution with an activity of $10(2) \ \mu \mathrm{Ci}$, which corresponds to $\sim 3\times10^{16}$ radium-226 atoms.  We made a laser ablation target by drying the radium solution on a 316 stainless steel mount which was installed in the vacuum system on a translation stage to position the target $\sim 15$ mm from the trap center, see Fig. \ref{fig:trap_levels_and_chain} (c).  

The radium ion fluoresces when near resonant light addresses the $S_{1/2} \rightarrow P_{1/2}$ transition at 468 nm and the $D_{3/2} \rightarrow P_{1/2}$ transition at 1079 nm, see Fig. \ref{fig:trap_levels_and_chain} (b).  The ion is laser cooled when the 468 nm laser is red detuned from the $S_{1/2} \rightarrow P_{1/2}$ transition.  Electronic branching from the $P_{1/2}$ state populates the $D_{3/2}$ state, which the 1079 nm light repumps back into the fluorescence cycle.  To prevent coherent dark states a magnetic field of a few gauss is applied \cite{Berkeland2002}.

The signals for our measurements are the 468 nm photons spontaneously emitted by radium ions.  These photons are focused onto a photomultipler tube (PMT), Hamamatsu H10682-210, whose output is sent to an integrated direct digital synthesizer and field-programmable gate array control and measurement system that can convert the PMT pulses to time-tagged photons \cite{Pruttivarasin2015}.  The same system synchronously controls the measurement sequences by driving acousto-optic modulators (AOMs) which set the amplitude and frequency offsets for the 468 and 1079 nm lasers.  The AOM extinction ratios are $\geq 60$ dB.  Measurement sequences, based on the techniques developed by Ramm \emph{et al.} \cite{Ramm2013} and Pruttivarasin \emph{et al.} \cite{Pruttivarasin2014a}, eliminate challenging systematics, such as AC Stark shifts, by addressing only one transition at a time.

\begin{figure}
    \centering
    \includegraphics[scale=1.0]{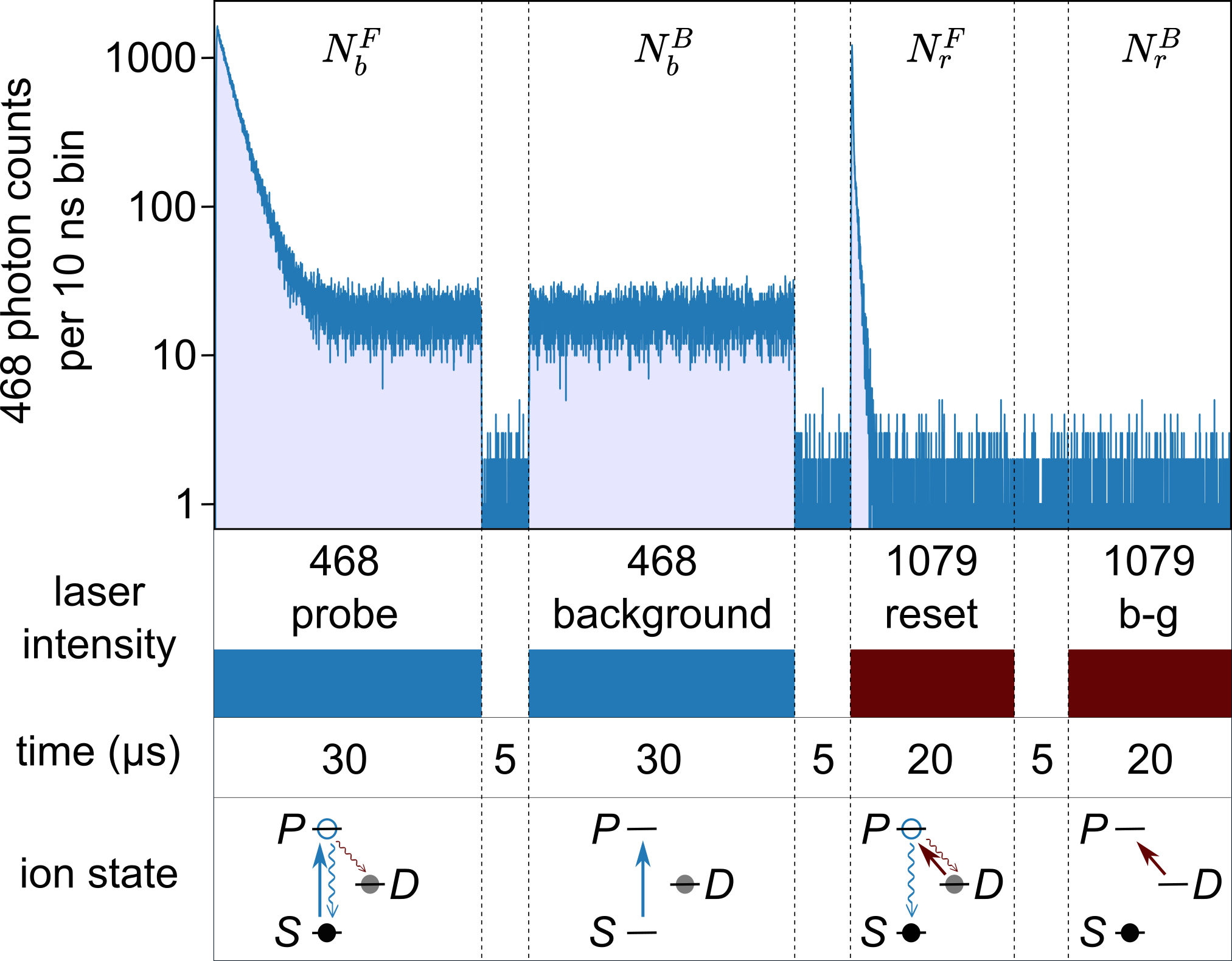}
    \caption{The total PMT counts during the $P_{1/2}$ branching fraction measurement are shown, along with the corresponding laser pulses and the measurement timing sequence.  The bottom panel shows the electronic population, applied optical fields, and expected decays.}
    \label{fig:branching_pulse_sequence}
\end{figure}

The branching fraction measurement sequence is summarized in Fig. \ref{fig:branching_pulse_sequence}.  Before each measurement, we Doppler cool the radium ion for 100 $\mu s$ and then optically pump population into the $S_{1/2}$ state by applying 1079 nm light for 20 $\mu s$.  After cooling and state preparation the electronic population is optically pumped to the $D_{3/2}$ state and collected 468 nm photons are time tagged, $N_b^F$.  The background scattered 468 nm light is then measured, $N_b^B$, while the ion is shelved in the $D_{3/2}$ state.  The population is then pumped back to the $S_{1/2}$ ground state with 1079 nm light, and if the single emitted 468 nm photon is collected it is time tagged, $N_r^F$, and a corresponding background, $N_r^B$, is recorded.  We subtract the respective backgrounds to determine the number of collected photons emitted by the radium ion, $N_b$ and $N_r$. From these counts we calculate the branching fraction to the ground state, $p = [N_b/(N_b+N_r)]$ \cite{Ramm2013}.  The measurement is repeated $11.5\times 10^6$ times in approximately one hour with a single radium ion. The raw photon counting results are $N_b^F = \num{359583}$, $N_b^B = \num{55297}$, $N_r^F = \num{31386}$, and $N_r^B = \num{1443}$, which yields a statistical branching fraction of $p=0.9104(5)$. The $N_r$ counts are also used to measure the imaging system detection efficiency (0.26\%).

The largest systematic uncertainty in the branching fraction measurement comes from residual birefringence of the imaging system and the Hanle effect. If the 468 and 1079 nm lasers are perfectly linearly polarized, the Hanle effect is not present, and an equal number of right- and left-handed circularly polarized photons will be collected \cite{Gallagher1963}. However, if either laser beam has a circularly polarized component, then there will be an imbalance in the right- and left-handed circularly polarized photons collected.  The imbalance will depend on the direction and magnitude of the applied magnetic field which sets the quantization axis. Residual birefringence of the imaging system may result in different detection efficiencies for the two circular polarizations, which in turn will shift the branching fraction measurement. The applied magnetic field is parallel to the 1079 nm laser, and both 1079 and 468 nm lasers are linearly polarized to suppress the Hanle effect. We set a limit on the uncertainty due to residual birefringence and the Hanle effect by reversing the applied magnetic field. The field reversal will flip the imbalance between right- and left-handed circularly polarized photons collected \cite{Zimmermann1975}, giving a different value for the branching fraction \cite{Ramm2013, Supplemental}. The measured branching fraction with the field reversed, $p = 0.9107(5)$, agrees with the original field configuration.  Therefore, uncertainty due to the combined effects of residual birefringence and the Hanle effect is at the level of the statistical uncertainty.


The systematic uncertainties and shifts due to other sources we considered are significantly less than those due to residual birefringence.  For shifts due to collisions we considered worst case scenarios.  For example, after shelving to the $D_{3/2}$ state, a collision could put the ion in an orbit that is dark to the 1079 nm pump pulse, and then a second collision could return the ion to the trap center where it emits a 468 nm photon during the 1079 nm background pulse. From our measurements we estimate the average collision rate to be less than 0.32 collision per second. PMT dead time (20 ns) results in both systematic error and shift.  We evaluate systematic shifts due to the finite lifetime of the $D_{3/2}$ state, the finite measurement time, and the finite extinction ratio of the AOMs using optical Bloch equations that describe the three-level system \cite{Ramm2013}. To solve the Bloch equations we use Rabi frequencies determined from fitting the spontaneous decays in $N_b^F$ and $N_r^F$, a theoretical $D_{3/2}$ state lifetime of 638(10) ms \cite{Pal2009}, and the $P_{1/2}$ branching fraction, $p = 0.9104$, from our statistical results. More details are included in the Supplemental Material \cite{Supplemental}.

\begin{figure}
    \centering
    \includegraphics[scale=1.0]{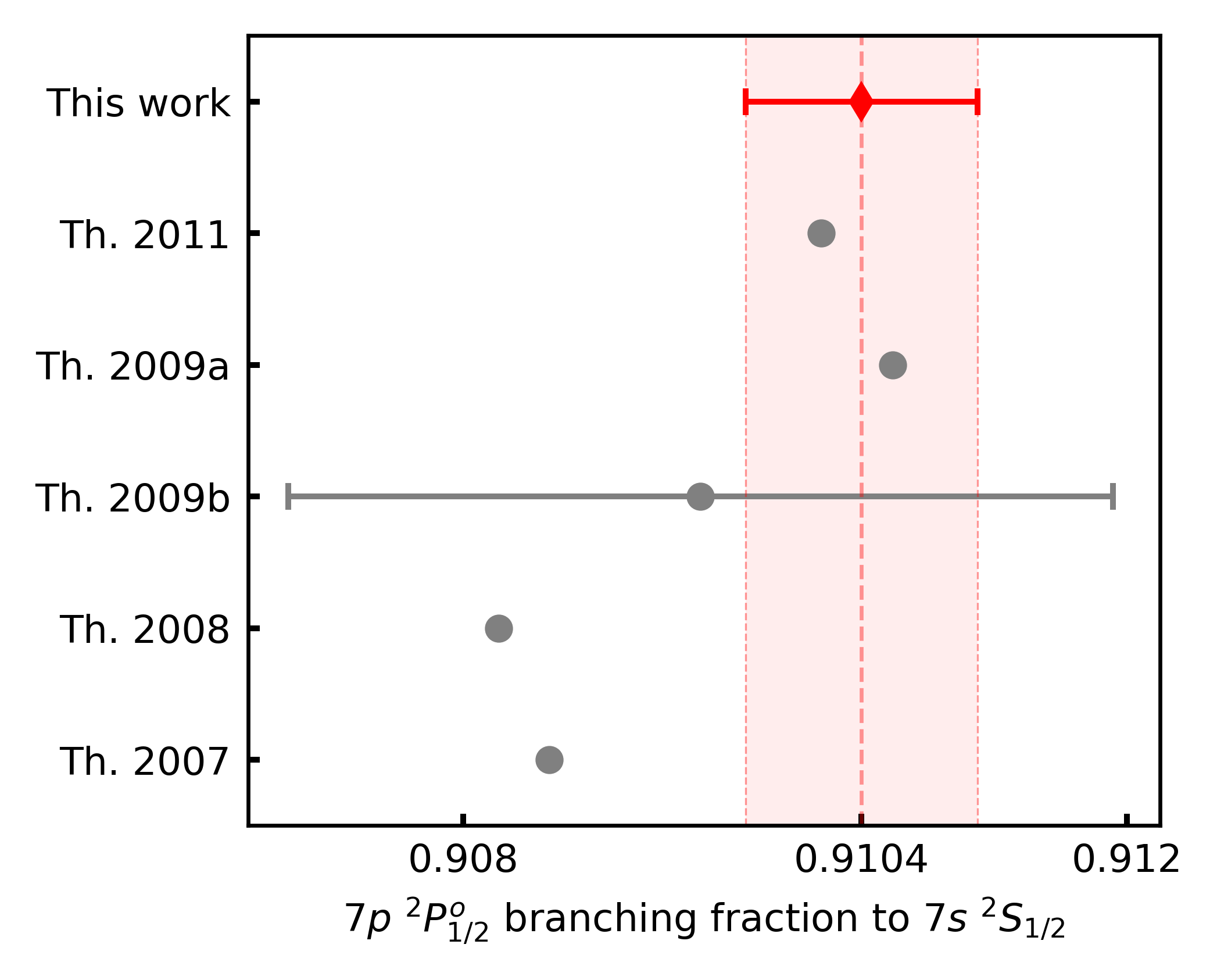}
    \caption{The measured branching fraction of the $^{226}\mathrm{Ra}^+$ $P_{1/2}$ to the $S_{1/2}$ (diamond), $p = 0.9104(7)$, compared to previous theoretical values (circles) where error bars are included when uncertainty is available.  The theoretical branching ratios are determined from reduced dipole matrix elements in the corresponding references: Th. 2011 \cite{Dzuba2011}, Th. 2009a \cite{Pal2009}, Th. 2009b \cite{Sahoo2009}, Th. 2008 \cite{Wansbeek2008}, Th. 2007 \cite{Sahoo2007} .}  
    \label{fig:branching_theory_comparison}
\end{figure}

The uncertainties and shifts for the branching fraction measurement are summarized in Table \ref{table:branching_fraction_systematics}.  When we add the uncertainties in quadrature the branching fraction to the ground state is $p = 0.9104(7)$, where systematic shifts do not contribute as their sum is far below the measurement uncertainty.  The measurement verifies theoretical techniques applied to this multielectron system that previously gave the only knowledge of the branching fraction \cite{Pal2009, Sahoo2009, Dzuba2011}, see Fig. \ref{fig:branching_theory_comparison}.  The  measurement can also be expressed as a ratio of the reduced dipole matrix elements between the $S_{1/2} \rightarrow P_{1/2}$ and $D_{3/2} \rightarrow P_{1/2}$ transitions, $m_{SP}/m_{PD}=0.912(4)$.

\begin{table} \centering
\caption{Uncertainties and shifts for the $P_{1/2}$ branching measurement.
\label{table:branching_fraction_systematics}}
\begin{ruledtabular}
\begin{tabular}{lcc}
    Source & Shift & Uncertainty  \\ 
    \midrule
    Statistical & ... & $5\times10^{-4}$ \\
    Birefringence & ... & $5\times10^{-4}$ \\
    Collisions &  ... & $<4\times10^{-5}$ \\
    PMT dead time & $3\times10^{-6}$ & $3\times10^{-6}$ \\
    $D_{3/2}$ state lifetime & $2\times10^{-7}$ & $2\times10^{-8}$ \\ 
    Measurement time & $5\times10^{-9}$ & $3\times10^{-7}$ \\ 
    AOM extinction ratio & ... & $5\times10^{-7}$ \\ 
    \bottomrule \\[-7pt]
    \bf{Total } & $3\times10^{-6}$ & $7\times10^{-4}$ \\ 
\end{tabular}
\end{ruledtabular}
\end{table}

The $S_{1/2}\rightarrow P_{1/2}$ transition of $\mathrm{Ra}^+$ is crucial to laser cooling and state detection.  We measure this transition's linewidth and center frequency with a linescan measurement \cite{Pruttivarasin2014a}.  From the linewidth we infer a lower limit on the $P_{1/2}$ state's lifetime of 7.3(1) ns, consistent with the theoretical value of 8.57(10) ns \cite{Sahoo2009}.  With tellurium vapor cell spectroscopy we determine the ${}^{226}\mathrm{Ra}^{+}$ $7s \ ^2S_{1/2}\rightarrow 7p \ ^2P_{1/2}^o$ transition frequency to be \num{640.09663}(6) THz. Our measurement agrees with a transition frequency of \num{640.096647}(23) THz, which was inferred from a measurement of this transition in $^{214}\mathrm{Ra}^+$ anchored to tellurium line 178 \cite{NunezPortela2014},  and separate $\mathrm{Ra}^+$ isotope shift measurements at the CERN ISOLDE facility \cite{Wendt1987}.

\begin{figure}
    \centering
    \includegraphics[scale=1.0]{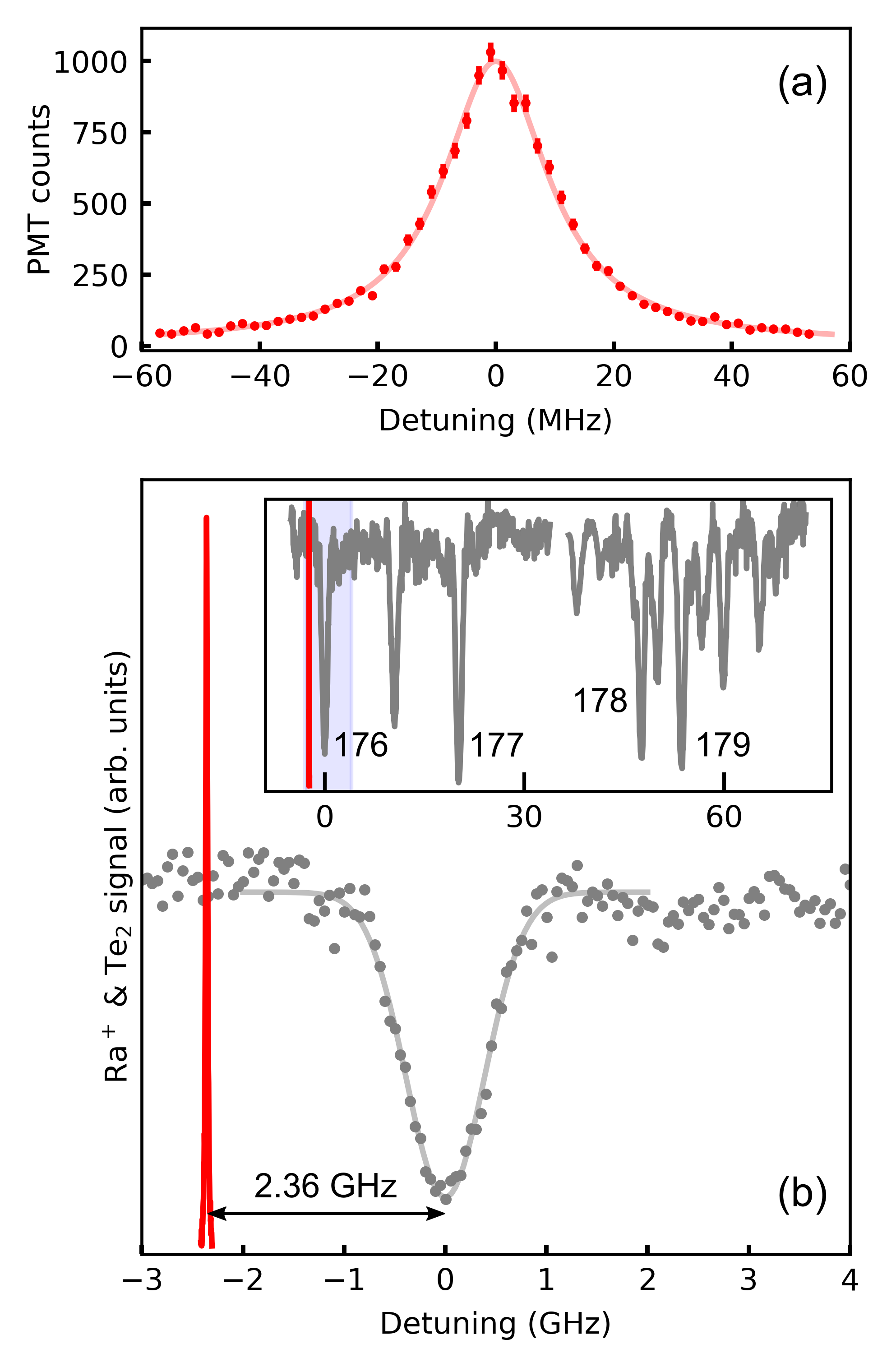}
    \caption{Collected photons from the $\mathrm{Ra}^+$ linescan measurement with a Lorentzian fit (a).  The tellurium absorption spectrum is plotted with the $\mathrm{Ra}^+$ linescan (b), where the tellurium data (grey) and $\mathrm{Ra}^+$ data (red) are scaled and offset to highlight the detuning between the transitions.  The inset of (b) shows tellurium lines in the vicinity of the $\mathrm{Ra}^+$ transition labelled with their atlas numbers, and the frequency span of the outset region highlighted in blue.}  
    \label{fig:linescan_spectroscopy}
\end{figure}

For the $\mathrm{Ra}^+$ linescan we trap and laser cool a chain of four radium ions. In the measurement sequence the ions are excited by a 468 nm laser probe pulse (2 $\mu$s), and then reset back to the ground state by a 1079 nm pulse (10 $\mu$s). The pulse sequence is repeated for different detunings, and before every ten sequences the ions are Doppler cooled for 500 $\mu$s and optically pumped to the electronic ground state.  We run the pulse sequence $2 \times 10^5$ times at each of 56 detunings set by an AOM with randomized measurement ordering.  The 468 nm laser is Pound-Drever-Hall locked to a Corning Ultra-Low Expansion (ULE) glass cavity sealed in a vacuum chamber with multiple layers of acoustic, seismic, and thermal isolation with active temperature stabilization \cite{Kumph2015}.  In order to determine the transition frequency by  a comparison with tellurium spectroscopy, the stabilized laser frequency during the measurement is recorded with a wave meter.   The photon counts for the measurement are plotted in Fig \ref{fig:linescan_spectroscopy}.

The nearest measured $^{130}\mathrm{Te}_{2}$ line to the radium transition is line 176 at \num{640.09899}(5) THz \cite{Cariou1980a}.  We measure the line in a 10 cm long tellurium vapor cell at 550$\degree$C by scanning a laser while recording the absorption on a photodetector and the frequency with a wave meter (High Finesse WS-8), see Fig. \ref{fig:linescan_spectroscopy} (b).  The line center is determined with a Gaussian fit and is then compared to the recorded frequency of the radium linescan to determine its detuning, -2.36 GHz.  The largest uncertainty in the frequency measurement is the 50 MHz uncertainty in the $\mathrm{Te}_{2}$ line \cite{Cariou1980a}.  There is a 10 MHz uncertainty contribution due to the wavemeter, which is determined with multiple linescans of the radium ion's $S_{1/2}\rightarrow P_{1/2}$ transition, and an additional 10 MHz uncertainty in the measured $\mathrm{Te}_{2}$ transition's center frequency.

We fit the photon counts of the $\mathrm{Ra}^+$ linescan to a Lorentzian and get a linewidth of 21.7(4) MHz, see Fig. \ref{fig:linescan_spectroscopy} (a).  The largest broadening contribution is likely micromotion Doppler broadening \cite{Berkeland1998}, which we estimate broadens the line by  $\sim 2.1$ MHz \cite{Supplemental}.

In this work we laser cooled trapped $\mathrm{Ra}^+$, an element where the most stable isotope, radium-226, has a 1600 year half-life, in a tabletop experiment ($<4$ L vacuum volume).  Laser cooling the trapped radium ions helped keep the ions well localized in the trap for $>12$ h at a time, enabling a precision measurement of the $P_{1/2}$ state's branching fraction.

This work opens the door to research with laser-cooled radium ions, including isotopes such as radium-225.  The low charge-to-mass ratio of $\mathrm{Ra}^+$ is well suited to sympathetic cooling of heavy atoms and large molecular ions.  Cold $\mathrm{Ra}^+$ could be used to make molecular ions such as $\mathrm{RaOH}^+$, and to sympathetically cool their motion and control their internal states with quantum logic spectroscopy \cite{Chou2017}.

We thank J. Roten and C. Schneider for technical assistance, and W. Campbell, A. Derevianko, D. Hucul, D. Patterson, A. Ransford, C. Schneider, T. Pruttivarasin, and A. Vutha for discussions and acknowledge support from the UC Office of the President
(MRP-19-601445).

\end{document}